\documentclass[twocolumn,
superscriptaddress,
 amsmath,amssymb,
aps,
prl,
]{revtex4}

\usepackage{graphicx}
\usepackage{dcolumn}
\usepackage{bm}

\usepackage{SIunits}

\usepackage{pict2e}

\usepackage{color}

\usepackage{ulem}

\newcommand{\red}[1]{\textcolor{black}{#1}}
\newcommand{\green}[1]{\textcolor{black}{#1}}

\newcommand{\dd}{\mathrm{d}}

\begin{document}

\title{
\red{Reducing} phonon-induced decoherence in \red{solid-state} single-photon sources with cavity quantum electrodynamics
}

\author{T. Grange}
\email{thomas.grange@neel.cnrs.fr}
\affiliation{Universit\'e Grenoble Alpes, F-38000 Grenoble, France}
\affiliation{CNRS, Institut N\'eel, “Nanophysique et Semiconducteurs” Group, F-38000 Grenoble, France}

\author{N. Somaschi}
\email{niccolo.somaschi@lpn.cnrs.fr}
\affiliation{Centre de Nanosciences et de Nanotechnologies, CNRS, Univ. Paris-Sud, UMR 9001, Universit\'e Paris-Saclay, C2N -- Marcoussis, 91460 Marcoussis, France}

\author{C. Ant\'on}
\affiliation{Centre de Nanosciences et de Nanotechnologies, CNRS, Univ. Paris-Sud, UMR 9001, Universit\'e Paris-Saclay, C2N -- Marcoussis, 91460 Marcoussis, France}

\author{L. De Santis}
\affiliation{Centre de Nanosciences et de Nanotechnologies, CNRS, Univ. Paris-Sud, UMR 9001, Universit\'e Paris-Saclay, C2N -- Marcoussis, 91460 Marcoussis, France}
\affiliation{Universit\'e Paris-Sud, Universit\'e Paris-Saclay, F-91405 Orsay, France}

\author{G. Coppola}
\affiliation{Centre de Nanosciences et de Nanotechnologies, CNRS, Univ. Paris-Sud, UMR 9001, Universit\'e Paris-Saclay, C2N -- Marcoussis, 91460 Marcoussis, France}

\author{V. Giesz}
\affiliation{Centre de Nanosciences et de Nanotechnologies, CNRS, Univ. Paris-Sud, UMR 9001, Universit\'e Paris-Saclay, C2N -- Marcoussis, 91460 Marcoussis, France}

\author{A. Lema\^itre}
\affiliation{Centre de Nanosciences et de Nanotechnologies, CNRS, Univ. Paris-Sud, UMR 9001, Universit\'e Paris-Saclay, C2N -- Marcoussis, 91460 Marcoussis, France}

\author{I. Sagnes}
\affiliation{Centre de Nanosciences et de Nanotechnologies, CNRS, Univ. Paris-Sud, UMR 9001, Universit\'e Paris-Saclay, C2N -- Marcoussis, 91460 Marcoussis, France}

\author{A. Auff\`eves}
\email{alexia.auffeves@neel.cnrs.fr}
\affiliation{Universit\'e Grenoble Alpes, F-38000 Grenoble, France}
\affiliation{CNRS, Institut N\'eel, “Nanophysique et Semiconducteurs” Group, F-38000 Grenoble, France}

\author{P. Senellart}
\email{pascale.senellart@lpn.cnrs.fr}
\affiliation{Centre de Nanosciences et de Nanotechnologies, CNRS, Univ. Paris-Sud, UMR 9001, Universit\'e Paris-Saclay, C2N -- Marcoussis, 91460 Marcoussis, France}
\affiliation{D\'epartement de Physique, Ecole Polytechnique, Universit\'e Paris-Saclay, F-91128 Palaiseau, France}

\date{\today}

\begin{abstract}

Solid-state emitters are excellent candidates for developing integrated sources of single photons. Yet, phonons degrade the photon indistinguishability both through pure dephasing of the zero-phonon line and through phonon-assisted emission. Here, we study theoretically and experimentally the indistinguishability of photons emitted by a semiconductor quantum dot in a microcavity as a function of temperature.
 \red{We show that
a large coupling to a high quality factor cavity can
simultaneously
 reduces the effect of  both phonon-induced sources of decoherence. It first limits the effect of pure dephasing on the zero phonon line with indistinguishabilities above 97\%  up to 18K. Moreover,  it efficiently redirects the phonon sidebands into the zero-phonon line and brings the indistinguishability of the full emission spectrum from 87\% (resp. 24\%) without cavity effect to more than 99\%  (resp $76\%$) at $0~$K (resp. $20~$K). We provide guidelines for optimal cavity designs that further minimize the phonon-induced decoherence.}

\end{abstract}

\maketitle

Indistinguishable single photons are the building blocks of optical quantum computation protocols and quantum networks \cite{aspuru2012photonic,he2016scalable,delteil2015generation}.
This has motivated great efforts to develop devices generating on-demand indistinguishable single photons, using solid-state emitters such as diamond color centres \cite{bernien2013heralded,sipahigil2014indistinguishable}, molecules \cite{lettow2010quantum} or semiconductor quantum dots (QDs) \cite{santori2002indistinguishable,gazzano2013bright,he2013demand,wei2014deterministic,somaschi2016near,ding2016demand}.
In QDs, understanding the extrinsic sources of decoherence such as spin and charge noise \cite{KuhlmannNoise} has recently enabled impressive progresses in the performances of these sources \cite{somaschi2016near,ding2016demand}. 
Yet, acoustic phonons generally remain an intrinsic and limiting source of dephasing. 

Indeed acoustic phonons are responsible for two kinds of dephasing processes.
First, acoustic phonons induce a rapid and partial decay of coherence \cite{borri01,krummheuer02,jakubczyk2016impact}. This non-Markovian dephasing dynamics is the time-domain counterpart of the emitter spectrum consisting of a sharp zero-phonon line (ZPL) sitting on top of a broad phonon sideband (PSB) \cite{besombes01, favero03,peter2004phonon}.
Second, acoustic phonons can assist virtual transitions towards higher energy levels, resulting in a Markovian pure dephasing of the ZPL \cite{muljarov04}.
Such effects impose two severe limitations to obtain indistinguishable photons: (i) to work at low temperatures, typically below 10~K for QDs and (ii) to use spectral postselection of the ZPL. Indeed, \green{even at zero temperature, phonon emission processes result in the presence of a PSB on the low energy side, fundamentally limiting the indistinguishability}. In practice, the indistinguishability has been measured to rapidly drop with temperature  even with spectral selection \cite{thoma2016exploring}, and without it remains further away from unity \cite{monniello2014indistinguishable}.

In typical self-assembled QDs, the emission fraction into the ZPL, $\eta_{\text{ZPL}}$, represents typically $90 \%$ of the emission at 4K, a fraction that rapidly drops with temperature. Moreover, the photons emitted in the PSB are essentially incoherent, due to their broadband nature with respect to the natural linewidth. 
In a Hong-Ou-Mandel (HOM) experiment, only the fraction $\eta_{\text{ZPL}}$ from each of the two photons gives rise to an interference. As a consequence, the indistinguishability (i.e. the degree of two-photon interference) decreases as $\sim \eta_{\text{ZPL}}^2$ \cite{kaer2014decoherence}.
To avoid this problem, a spectral filtering of the collected emission spectrum, corresponding to a spectral post-selection,  is generally applied to spectrally select the ZPL while suppressing the phonon sideband emission \cite{gazzano2013bright,he2013demand,wei2014deterministic,somaschi2016near,thoma2016exploring, loredo2016scalable,ding2016demand}.
Yet, such postselection of the emission constitutes a fundamental limitation since it automatically reduces the efficiency of the single-photon source. Besides, the finite efficiency of the spectral filter further decreases the efficiency, shifting away from the sought deterministic characteristic i.e. the generation of exactly one indistinguishable photon per pulse.
It is hence of prime interest to develop solid-state sources of highly indistinguishable photons that do not require postselection.

Here we study experimentally and theoretically the indistinguishability of photons emitted by QDs inserted in micropillar cavities as a function of temperature. We demonstrate that cavity-quantum electrodynamics (cavity-QED) effects can \red{be used to efficiently reduce} the different phonon-induced dephasing effects.
Strong Purcell enhancement of the radiative emission rate has for long been thought as a promising strategy to reduce the influence of pure dephasing, but experimental demonstrations have \red{been limited by the fact} that increasing temperature also modifies the QD-cavity coupling \cite{varoutsis2005restoration,unsleber2015two}. Here, we use an electrical tuning of the QD transition to keep the ZPL resonant to the cavity mode \red{ and observe indistinguishabilities of the ZPL above 97\% up to 18K. }
\red{We} also show that the coupling to a high quality factor cavity  efficiently redirects the PSB emission into the ZPL, leading to a \red{significant} increase of $\eta_{\text{ZPL}}$. \red{At $0~$K (resp. $20~$K), this redirection allows bringing the indistinguishability from 87\% ( resp. 24\%) - limited by phonon emission processes - to  more than 99\% (resp $76\%$). Our theoretical study,   based on  non-equilibrium Green's functions calculations, predicts a further improvement of the indistinguishability of postselection-free  single photons by  reducing the cavity linewidth for a fixed Purcell factor.}

\begin{figure}
\begin{centering}
\includegraphics[width=0.5\textwidth]{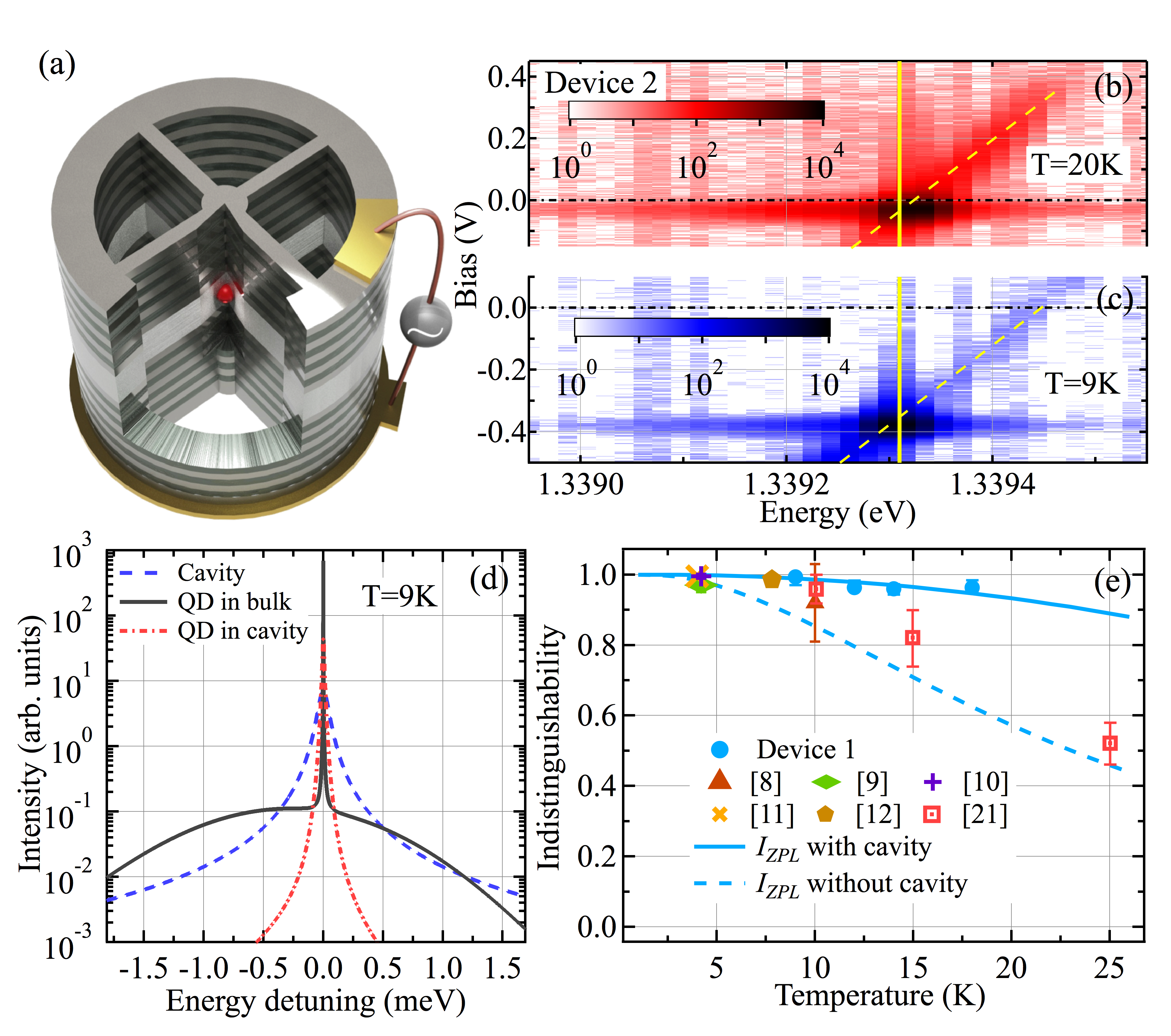} 
\end{centering}
\caption{(Color online). 
(a) Schematic of the device. (b,c) Resonant PL maps as function of energy and bias voltage on Device 2 at temperatures 20 and 9 K, respectively. The cavity and QD spectral positions as function of the voltage are indicated with full and dashed yellow lines, respectively. (d) Comparison of the calculated emission spectra of a QD in a bulk photonic environment (black solid line) and coupled resonantly to a cavity (red dashed-dotted line) at 9K. The cavity spectrum is also indicated (blue dashed line). (e) Indistinguishability of the ZPL as a function of temperature. The measurements (Device 1) are shown in blue circles. Calculations for the Device~1 (solid line) and without cavity-QED effects (dashed line) are shown. \red{We also indicate measurements reporting high indistinguishability of the ZPL as well as the temperature dependence recently reported in ref.   \cite{thoma2016exploring}  in the absence of Purcell effect (red squares).} 
}
\label{Fig1}
\end{figure}

The QD-cavity devices are obtained through a deterministic positioning of the QD in the cavity using the in-situ lithography technique \cite{dousse2008}. The micropillar cavity is electrically contacted with four ridges to a surrounding circular structure where the electrical connection is performed [see sketch in Fig. \ref{Fig1}(a)]. The applied bias voltage allows for an accurate control on the detuning of the QD-cavity resonance \cite{Nowak:2014aa,somaschi2016near} by means of the confined Stark effect. In this work, we study two different QD-cavity devices, dubbed Device 1 and 2. 
The cavity linewidth is 90 $\mu$eV (110 $\mu$eV) for the Device 1 (2). In both cases the electronic transition corresponds to a neutral exciton constituted by two orthogonal and linear dipoles showing a fine structure splitting of $\Delta_\mathrm{FSS}= 3$ (resp. $10$) $\mu$eV for Device 1 (resp. 2).  The cavities  are excited with a resonant laser using a standard confocal resonant fluorescence setup in a  cross-polarization configuration to remove the resonant laser light \cite{he2013demand,somaschi2016near,giesz2016coherent}.
Such strict resonant excitation suppresses the effect of time jitter in the exciton creation process \cite{kiraz2004quantum}, thus isolating the effect of phonon-induced decoherence on indistinguishability.

The sample temperature is controlled between  9 to 20K in a closed-circuit gaz exchange helium cryostat.
Measuring the effect of cavity-QED on the photon indistinguishability when changing the temperature \red{requires an additional knob to keep the QD-cavity detuning constant since the QD transition shifts faster with temperature than the cavity resonance} \cite{varoutsis2005restoration,unsleber2015two}.  Here, the QD-cavity resonance is maintained \red{with} temperature  through the confined Stark effect.
Figures \ref{Fig1}(b,c) show two resonance fluorescence maps as function of the applied bias and energy at 20 K and 9 K, respectively, for Device 2: a continuous-wave laser resonantly excites one of the cavity modes (energy indicated with the vertical solid line)  and the crossed-polarized resonant fluorescence is collected. The voltage control allows tuning the QD resonance (dashed line) and reaching the QD-cavity resonance at 20 K and 9 K  for an applied bias of -26 and -376 mV, respectively. Note that above 20-23 K, depending on the device,  the resonance  is reached for a positive bias for which a non-negligible current is observed, changing the measurement conditions.

Figure \ref{Fig1}(c) evidences an asymmetric \red{calculated} spectrum at resonance at 9K, arising from the PSB emission \cite{besombes01, favero03,peter2004phonon}. This strongly non-Lorentzian shape is reproduced using a non-Markovian modeling of the phonon bath \cite{wilson2002quantum,machnikowski2004resonant,vagov2007nonmonotonic,ramsay2010damping,mccutcheon2010quantum,kaer2010non,hohenester2010cavity,roy2011phonon,hughes2011influence,glassl2011influence,debnath2012chirped,valente2014frequency,roy2015spontaneous,portalupi2015bright,manson2016polaron,nazir2016modelling}.  Accounting for the measured spectrum at 9~K at the QD-cavity resonance allows extracting the potential deformation constant of $D=14$~eV for the excitonic transition (see Supplemental Material \cite{supp}).
To highlight the influence of the cavity on the spectrum, the exciton spectrum are compared in Fig.~\ref{Fig1}(d) for a bulk-like emission (i.e. without cavity) and for the QD in cavity using a logarithmic scale. In the bulk case, the sharp ZPL (width of $0.7 \micro$eV) sits on meV-broad PSBs. When coupling the QD to the cavity, two important changes are observed: (i) the Purcell enhancement of the radiative emission broadens the ZPL, and (ii) the phonon PSBs that fall outside the cavity spectrum (dashed blue line) have their emission strongly suppressed. The PSB emission is hence effectively redirected towards the ZPL: the reduction of the efficiency of phonon-assisted emission together with the increase of the ZPL one strongly modifies the ratio between these two components of the spectrum. The fraction emitted in the ZPL at 9~K is calculated to increase from $\eta_{\text{ZPL}} = 0.81$ in bulk to $\eta_{\text{ZPL}} = 0.98$ in the cavity.
\red{The  spectra measured at various temperatures are found to be in good agreement with these theoretical expectations (see Fig.~S6 in Supplemental Material~\cite{supp}).} 
\red{Both cavity-QED effects significantly influence  the indistinguishability}.

We measure the two-photon coalescence in a fiber based HOM setup. The excitation consists on resonant pulses of 15 ps temporal length with a repetition rate of 82 MHz and a pump power corresponding to the $\pi$-pulse of the QD transition, in resonance with the cavity mode for every temperature. The single-photon emission is sent to a fiber based Mach-Zehnder interferometer, with a relative temporal delay between the two arms of 12.2~ns corresponding to the repetition rate of the laser. The output signals from the HOM interferometer are temporally correlated to reconstruct the two-photon coincidence histogram which renders the value of the indistinguishability \cite{supp}.

Using an etalon to filter the PSB (spectral width of 10~$\micro$eV, transmission efficiency of 70~\%), we first study the indistinguishability of photons emitted by the ZPL as a function of temperature (Fig. ~\ref{Fig1}(e)).
The indistinguishably of the ZPL hardly decreases from $0.993\substack{+0.006 \\ -0.024}$ at 9~K to $0.973\substack{+0.021 \\ -0.010}$ at 18~K. \red{Indistinguishabilities above $90\%$ of the ZPL have been reproducibly reported in the  4-10 K range (see symbols in Fig. ~\ref{Fig1}(e)) \cite{gazzano2013bright,he2013demand, wei2014deterministic,somaschi2016near,ding2016demand,thoma2016exploring}.
At higher temperatures,
the indistinguishability of the ZPL has been reported to 
decrease rapidly. This trend was observed both  on QD  in microlenses \cite{thoma2016exploring} (red squares in \ref{Fig1}(e)) and on  QDs in photonic waveguides where no acceleration of spontaneous emission is implemented. The  robust two-photon interferences with temperature observed here } can be attributed to the strong  Purcell effect  provided by the cavity coupling \cite{varoutsis2005restoration,ates2009post,unsleber2015two}. Indeed, for a two-level system with a Markovian dynamics, the indistinguishability is given by \cite{bylander2003interference, grange2015cavity}
\begin{equation}
I_{\text{ZPL}}= \gamma / (\gamma + \gamma^*),
\label{IZPL}
\end{equation}
where $\gamma=\hbar/T_1$ is the natural linewidth ($T_1$ being the exciton lifetime) and $\gamma^*= 2\hbar /T_2^*$ is the additional broadening of the ZPL due to pure dephasing.
Increasing the temperature degrades the indistinguishability of the emitted single photons by reducing the pure dephasing time $T_2^*$. Yet the Purcell effect has a positive impact on the indistinguishability by accelerating the single-photon radiative lifetime ($T_1$).
Indeed, by placing the QD in resonance with the microcavity, the radiative rate $\gamma$ increases from $\gamma = \gamma_0$ to $\gamma=(1+F_{\text{eff}})\gamma_0$, where $F_{\text{eff}}$ is the effective Purcell factor describing  the reduction from  the nominal Purcell factor ($F_P=24$ for Device 1) due to the presence of the PSBs  \cite{portalupi2015bright, supp}: $F_{\text{eff}}$ decreases with increasing temperature, from 15 at 0~K to 11 at 20~K.
As negligible pure dephasing is observed at 8K, the broadening of the ZPL is attributed to the phonon-assisted virtual transitions with the higher excitonic states.
Such mechanism involves a scattering from a thermal acoustic phonons into another mode, so that the resulting temperature dependence can be approximated by $\gamma^*(T) = \alpha n(\epsilon_p) [n(\epsilon_p)+1]$, where $n(\epsilon_p)$ is the Bose factor at the wavelength of maximally coupled acoustic phonons $\epsilon_p$ \cite{grange2009decoherence}.
We take $\alpha=0.1~\micro$eV and $\epsilon_p=1~$meV to account for the observed dependence of the ZPL indistinguishability.
Figure \ref{Fig1}(e) also shows the corresponding indistinguishability without any Purcell effect as $I_{\text{ZPL}}^{0} = \gamma_0 / (\gamma_0 + \gamma^*)$ (dashed line), evidencing the strong  enhancement of the ZPL indistinguishability induced by the large Purcell effect in Device~1.

\begin{figure}
\begin{centering}
\includegraphics[width=0.45\textwidth]{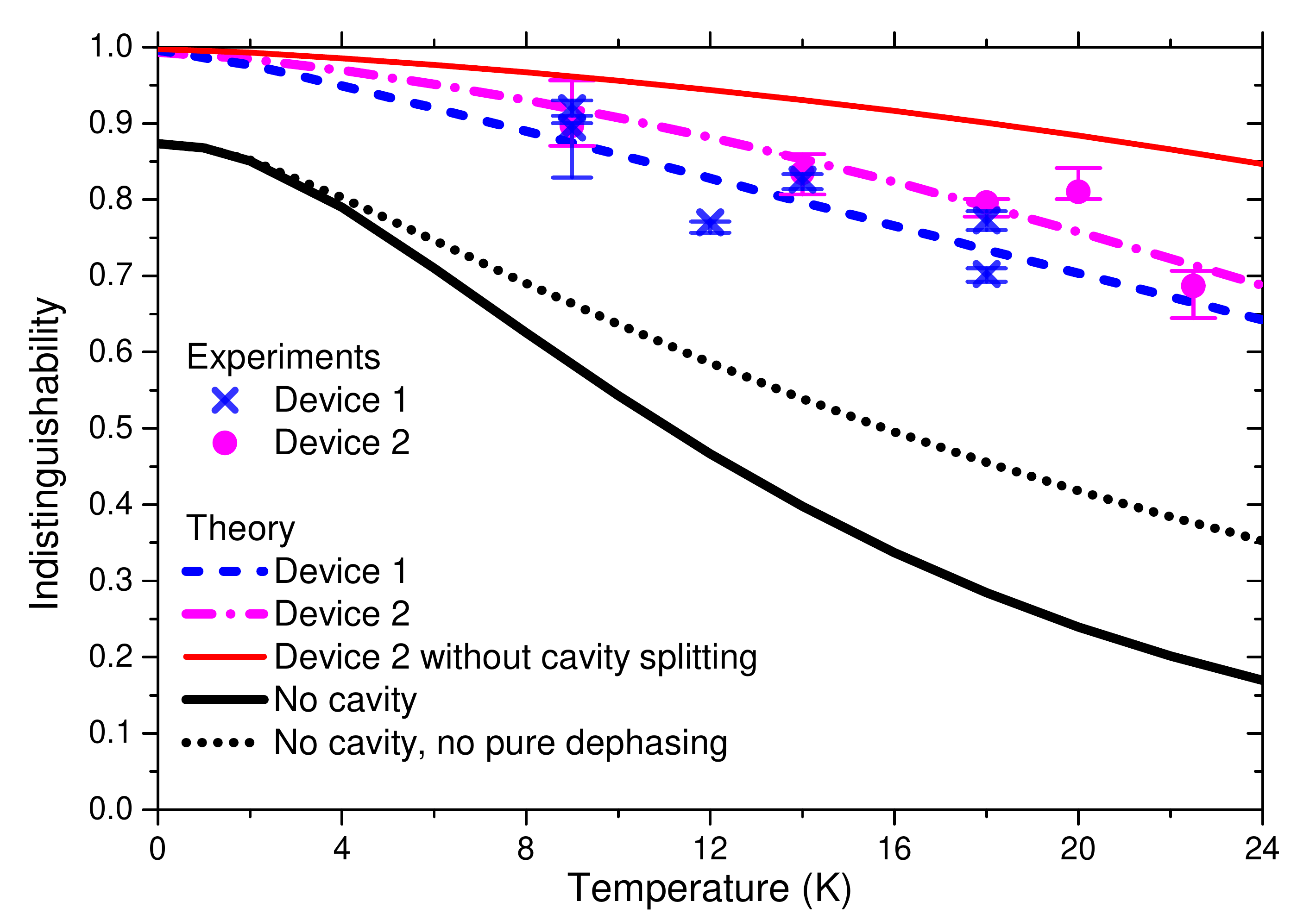} 
\end{centering}
\caption{(Color online).
Indistinguishability of the non-postselected photons as a function of temperature. The measurements are shown in symbols for the two different QD-cavity devices.
 Calculations accounting for the coupling to the cavity of Device 1 (Device 2) are indicated in dashed (dashed-dotted) lines.
 \red{Calculations in red thin solid line predicts the indistinguishability in the case where the cavity splitting is suppressed (the other parameters being identical to the one of Device~1).}
 Calculations in the absence of cavity are also shown (black solid line). The black dotted line represents the calculations in bulk in the absence of pure dephasing of the ZPL.
}
\label{IvsT}
\end{figure}

\red{In the following we study the photon indistinguishability without any spectral postselection. }
For a QD in bulk, the \red{calculated} indistinguishability is  shown (solid line) as a function of temperature without spectral postselection and accounting for the pure dephasing measured in Fig.~\ref{Fig1}(e). It  drops to 0.24 as temperature reaches 20 K.
This has been recently confirmed by similar measurements on QDs subject to negligible cavity-QED effects \cite{reigue2016probing}.
Even in the absence of pure dephasing (dotted line), the indistinguishability would fall to 0.42 at 20~K, owing to the rapid reduction of the ZPL fraction ($\eta_{\text{ZPL}}=0.64$ at 20~K)  (see Fig.~S5 in Supplemental Material~\cite{supp}) with increasing temperature. 
\red{To calculate the indistinguishability in the presence of cavity-QED}, the coupling to the non-Markovian phonon bath and to the cavity mode are simultaneously accounted for in the calculation of photon correlation \cite{kaer2013microscopic,kaer2013role,kaer2014decoherence,mccutcheon2016optical}.
We use a recently developed non-equilibrium Green's functions approach \cite{hornecker2016influence}. Its accuracy has been shown by the excellent agreement with the exact diagonalization reported at zero temperature by Kaer et al.  \cite{kaer2013microscopic}. Yet, in contrast, our model is numerically tractable at finite temperatures.
\red{The calculated indistinguishability} is shown in Fig.~\ref{IvsT} for the two different devices. The same QD parameters for PSBs and pure dephasing are considered for the two devices \cite{supp}. A strong enhancement of indistinguishability with respect to the bulk is predicted, due to the combination of (i) the Purcell effect overcoming pure dephasing of the ZPL [as already discussed in Fig~\ref{Fig1}(e)], and (ii) the cavity redirection of the PSB emission into the ZPL  (the relative contribution of these two effects is shown in Supplemental Material \cite{supp}). \green{Such redirection  appears first  at 0K where a close to unity indistinguishability is calculated with cavity-QED when it should be limited to $\sim0.87$ due to phonon emission processes. Considering finite temperatures, }the non-postselected indistinguishability is expected to decrease from $\sim0.92$ ($\sim0.89$) at 9~K to $\sim0.74$ ($\sim 0.79$) at 18~K for Device 1 (Device 2). \red{The measured indistinguishabilities} are also shown as a function of temperature for Device 1 (crosses) along with Device 2 (circles),  using voltage to recover the QD-cavity resonance at each temperature. Note that the values are corrected from a residual laser background to extract the emitted photon indistinguishability (see Supplemental Material \cite{supp}). \red{A very good agreement is found between the calculated and measured indistinguishabilities.}
Note that the dispersion in the measured points mainly arises from the $\pm 5 \micro$eV uncertainty on the QD-cavity detuning, which influences the indistinguishability \cite{supp}. Moreover device~1 does not provide a higher indistinguishability than Device~2 in spite of a much larger nominal Purcell factor (respectively 24 and 8). This is due to a larger residual cavity mode splitting which results in a larger detuning between the exciton and the monitored cavity mode under the crossed polarization resonant excitation scheme adopted here \cite{supp}. Such limitation would vanish under side excitation, and we plot the expected value for a negligible cavity mode splitting, where the indistinguishability at 20~K is predicted to further increase from 0.70 to 0.88 for Device~1, as shown in thin solid line in Fig.~\ref{IvsT}. 

As discussed below, such \red{enhancement} of indistinguishability with respect to a bulk-like photonic environment \cite{monniello2014indistinguishable} \red{results from} the small linewidth of the cavities with respect to the PSBs width. Indeed, as the PSBs that fall outside the cavity spectrum are efficiently redirected into the ZPL [Fig.~\ref{Fig1}(d)], a large fraction of incoherent emission is transformed into a coherent one. This second cavity-QED effect differs fundamentally from a filtering effect, for which PSB emission fraction would be simply lost: the overall CQED effect not only increases the photon indistinguishability, but also increases the overall source efficiency.   \red{Note that cavity-QED funneling into the ZPL was recently demonstrated for NV centers \cite{johnson2015tunable,NV2}, but with no measure yet of the full spectrum indistinguishability.}

\begin{figure}
\begin{centering}
\includegraphics[width=0.45\textwidth]{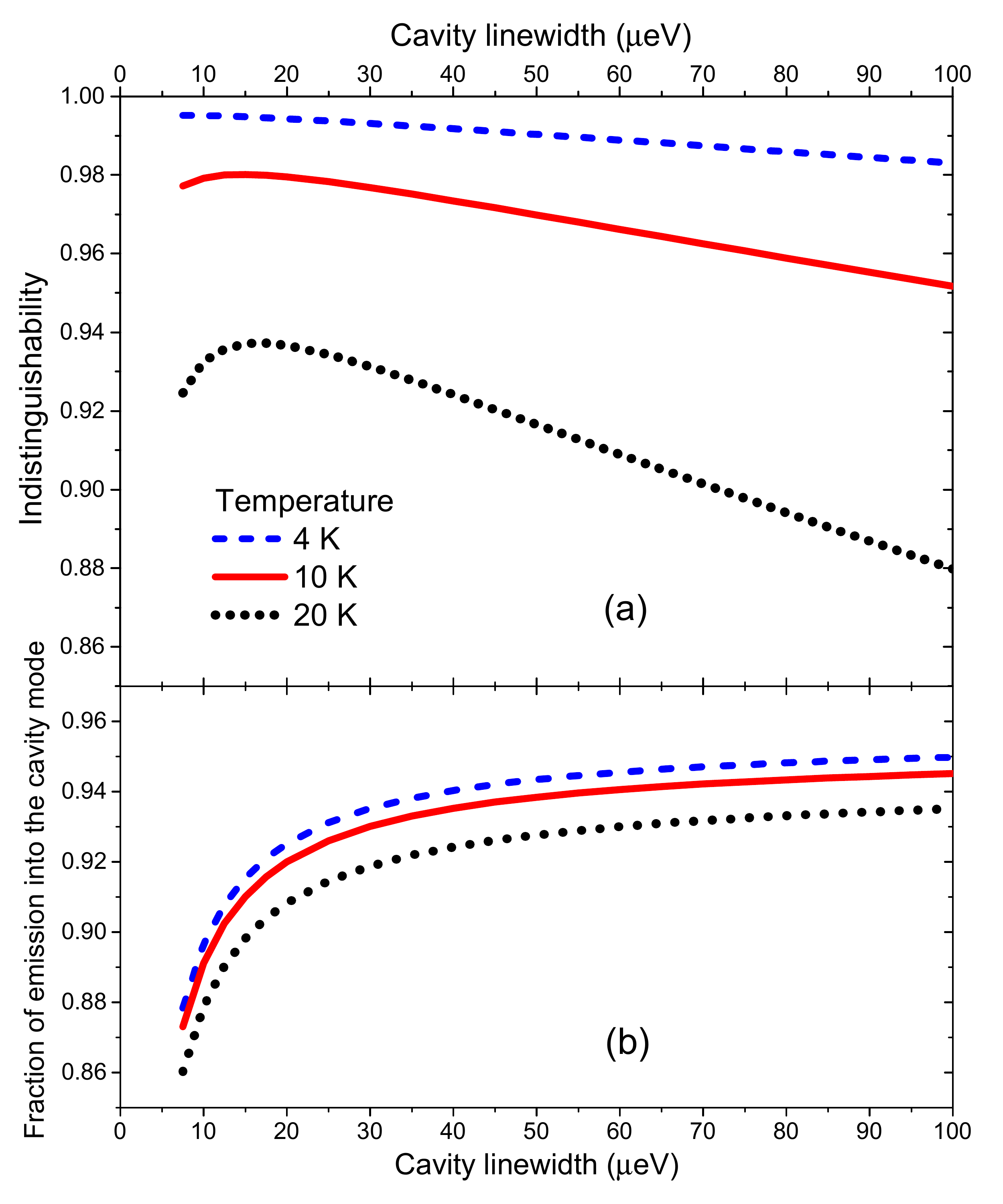} 
\end{centering}
\caption{(Color online). 
(a) Calculated indistinguishability of the non-postselected photons as a function of the cavity linewidth for a fixed nominal Purcell factor of $F=24$, for temperatures of 4, 10, and 20~K. (b) Corresponding fraction of the emission into the cavity mode for the same three temperatures. The QD and the cavity mode are assumed to be on resonance.
}
\label{Ivskappa}
\end{figure}

As shown above, a high Purcell factor \red{allows reducing}  the effect of ZPL pure dephasing. However, as shown by Kaer et al \cite{kaer2013microscopic,kaer2014decoherence}, \red{reaching} the QD-cavity strong coupling regime results in a decrease of indistinguishability
\red{due to phonon-assisted scattering processes between the dressed polariton states}.
Here instead, we propose to reduce the cavity linewidth to enhance the effect of the PSBs being redirected into the ZPL.
In Fig.~\ref{Ivskappa}(a), we show the calculated indistinguishability as a function of the cavity linewidth for a fixed nominal Purcell factor of 24 (as reported in Device~1). The QD-cavity detuning is set to its optimal zero value. 
Interestingly, the indistinguishability is found to further increase when the cavity linewidth is reduced below the cavity linewidths of our devices ($\sim100 \micro$eV).
 This effect is all the more important as the temperature increases, since  ($1-\eta_{\text{ZPL}} $), the nominal fraction of emission in the PSB, increases.
The indistinguishability is hence predicted to reach a value of 0.94 at 20K, and can be increased up to 0.995 at 4K.
 At very small cavity linewidths, the indistinguishability reaches a maximum before decreasing. Indeed, the strong QD-cavity coupling regime is reached for $\kappa \lesssim 15 \micro$eV and Eq.~\ref{IZPL} is then no longer valid \cite{grange2015cavity}.
Figure~\ref{Ivskappa}(b) shows the corresponding fraction of emission into the cavity mode.
Except for very small linewidths below $\sim 20 \micro$eV, the probability of emission into the cavity mode stays almost constant.
Hence, the reduction of the cavity linewidth enables enhancing the indistinguishability without decreasing significantly the brightness.
Experimentally, this can be realized by changing the cavity length of the micropillars from a single wavelength $\lambda$ to multiple $\lambda$ wavelengths, while keeping the same Bragg mirrors and hence almost the same Purcell factor (see Supplemental Material Fig.~S10 \cite{supp} for the influence of the Purcell factor).
This strategy can also be implemented in other cavity-QED systems where low cavity linewidths and large Purcell effects can be combined, such as
open-access cavities \cite{muller2009coupling,miguel2013cavity,greuter2014small} or photonic crystal cavities \cite{madsen2014efficient,kalliakos2014plane,lodahl2015interfacing,muller2016self}.

Our findings apply to a large variety of solid-state single-photon emitters suffering from phonon-induced decoherence.  Various approaches are explored to obtain bright sources of indistinguishable photons from solid-state emitters, some based on broadband spontaneous emission suppression \cite{bleuse2011inhibition,arcari2014near}, others, like here, based on acceleration of spontaneous emission into a narrowband mode.  We have shown that the latter approach, not only allows for high extraction efficiency, but for \red{reducing the effect of both phonon decoherence processes}. Indistinguishable photons from the ZPL are shown to be \red{less sensitive} to thermally-activated pure dephasing, thanks to the large Purcell acceleration of photon emission. Concurrently, indistinguishable photon generation without spectral postselection can be obtained  \red{at low temperature with high quality factor cavities by effectively reducing the phonon assisted emission processes}

\red{\textit{Note added in revision.} After our original submission, a related theoretical investigation was reported by Iles-Smith et al \cite{iles2016phonon}.}

This work was partially supported by the European Union's Seventh Framework Programme (FP7) under Grant Agreement No. 618078 (WASPS), the ERC Starting Grant No. 277885 QD-CQED, the French Agence Nationale pour la Recherche (grant ANR QDOM, SPIQE and USSEPP) the French RENATECH network, the Marie Curie individual fellowship SQUAPH, a public grant overseen by the French National Research Agency (ANR) as part of the "Investissements d'Avenir" program (Labex NanoSaclay, reference: ANR-10-LABX-0035), and the iXcore foundation.

\bibliographystyle{apsrev}

\pagebreak
\widetext
\begin{center}
\textbf{\large Supplemental Material: Reducing phonon-induced decoherence in solid-state single-photon sources with cavity quantum electrodynamics}
\medskip

T. Grange,$^{1, 2, *}$ N. Somaschi,$^{3, \dagger}$  C. Ant\'on,$^{3}$ L. De Santis,$^{3, 4}$ G. Coppola,$^{3}$ V. Giesz,$^{3}$ A. Lema\^itre,$^{3}$ I. Sagnes,$^{3}$ A. Auff\'eves,$^{1, 2,\ddag}$ and P. Senellart$^{3, 5,\S}$

\medskip

\textit{$^1$ Universit\'e Grenoble Alpes, F-38000 Grenoble, France\\
$^2$ CNRS, Institut N\'eel, Nanophysique et Semiconducteurs Group, F-38000 Grenoble, France\\
$^3$ Centre de Nanosciences et de Nanotechnologies, CNRS, Univ. Paris-Sud, Universit\'e Paris-Saclay, C2N -- Marcoussis, 91460 Marcoussis, France\\
$^4$ Universit\'e Paris-Sud, Universit\'e Paris-Saclay, F-91405 Orsay, France\\
$^5$ D\'epartement de Physique, Ecole Polytechnique, Universit\'e Paris-Saclay, F-91128 Palaiseau, France}
\end{center}

\setcounter{equation}{0}
\setcounter{figure}{0}
\setcounter{table}{0}
\setcounter{page}{1}
\makeatletter
\renewcommand{\theequation}{S\arabic{equation}}
\renewcommand{\thefigure}{S\arabic{figure}}
\renewcommand{\bibnumfmt}[1]{[S#1]}
\renewcommand{\citenumfont}[1]{S#1}

\section{Section 1: Fabrication of QD-cavity single-photon sources}

The sample was grown by molecular beam epitaxy on a substrate of $n$-doped (100) GaAs substrate. The planar cavity structure consists of: (i) a bottom Bragg mirror (n-doped) with 30 pairs made of $\lambda/4$ layers of GaAs and Al$_{0.95}$Ga$_{0.05}$As (thicknesses of 68 nm and 78 nm), (ii) a $\lambda$-GaAs cavity of a thickness of 274 nm embedding a layer self assembled InGaAs QDs, (iii) a top Bragg mirror (\textit{p}-doped) with 20 $\lambda/4$ layers. The planar microcavity  was processed using the in situ lithography technique to deterministically couple single QD to micropillar cavities in position and energy, with accuracies in spatial and energy matching of 50 nm and 0.5 meV \cite{dousse2008}. A bottom $n$-contact is deposited on the back of the sample while the top \textit{p-}contact is realized on a large frame to which the pillar is connected through one-dimensional wires. 

\section{Section 2: Experimental setup}

\begin{figure}
\begin{centering}
\includegraphics[width=0.8\textwidth]{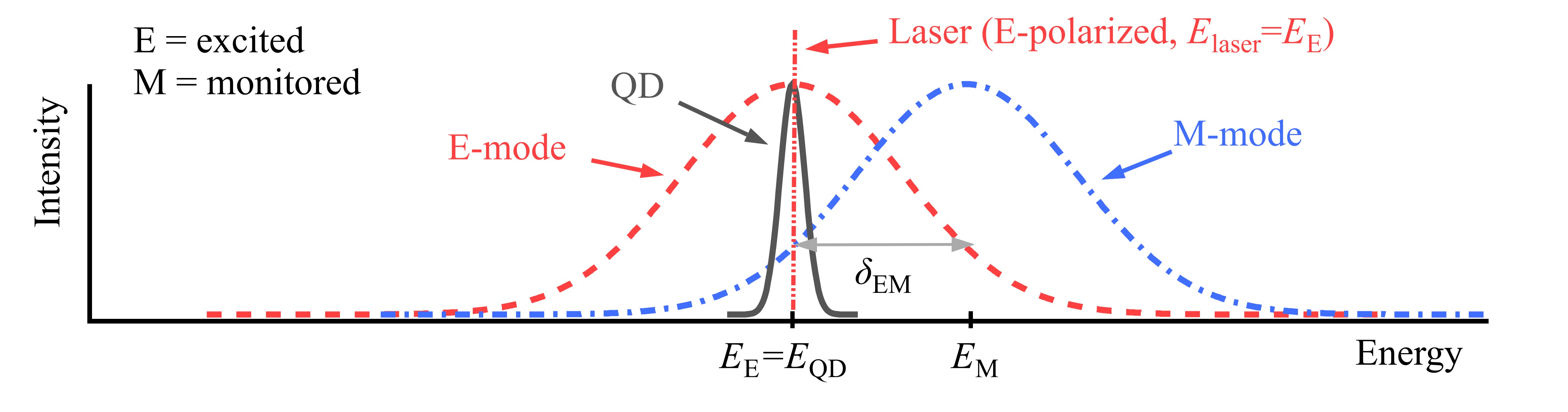} 
\end{centering}
\caption{Resonance fluorescence excitation scheme of the QD-cavity device: the QD is electrically tuned in resonance with the excited mode (E-mode) and the laser; the laser polarization is set in parallel to the one of the E-mode. The collection of the QD resonance fluorescence is done through the monitored mode (M-mode), orthogonally polarized to the E-mode. The energy splitting between the two modes is given by $\delta_{\textrm{EM}}$. According to the energy axis, the sketched excitation scheme corresponds to the Device 1, where $V$-mode (lower energy) is excited and $H$-mode (higher energy) is collected.}
\label{scheme_excitation}
\end{figure}

The optical excitation is performed with a resonant linearly polarized laser (continuous-wave or pulsed), which is focused on each device by a microscope objective after passing through a polarizing beamsplitter (PBS) and a half waveplate. The same objective is used to collect the QD resonance fluorescence coupled to a single mode fiber in a confocal geometry. A small anisotropy of the cavities leads to two linearly polarized cavity modes with a energy splitting of 80 (resp. 30) $\mu$eV for Device 1 (resp. 2). The laser polarization is aligned along one of the two orthogonal polarizations of the cavity axes ($V$/$H$ mode for Device 1/2), dubbed excited-mode (E-mode, see Fig. \ref{scheme_excitation}), and the laser energy is tuned to the corresponding one of the E-mode. 

In the devices under study, the electronic transition corresponds to a neutral exciton (see the QD line in the sketch of Fig. \ref{scheme_excitation}), with two orthogonal and linear dipoles showing a fine structure splitting of $\Delta_\mathrm{FSS}= 3$ (resp. $10$) $\mu$eV for Device 1 (reps. 2). The exciton transition is brought in resonance to the E-mode using the applied bias. As the cavity axes differs from the exciton ones, the excitation polarized along a cavity axis excites a coherent superposition of the two-exciton states that temporally evolve toward the orthogonal polarization. The resonant fluorescence is collected along this perpendicular polarization mode ($H$/$V$ mode for Device 1/2), labelled the monitored cavity mode (M-mode) in Fig. \ref{scheme_excitation}. For more details see reference \cite{giesz2016coherent}.

\section{Section 3: Analysis of two-photon correlations}

\begin{figure}
\begin{centering}
\includegraphics[width=0.7\textwidth]{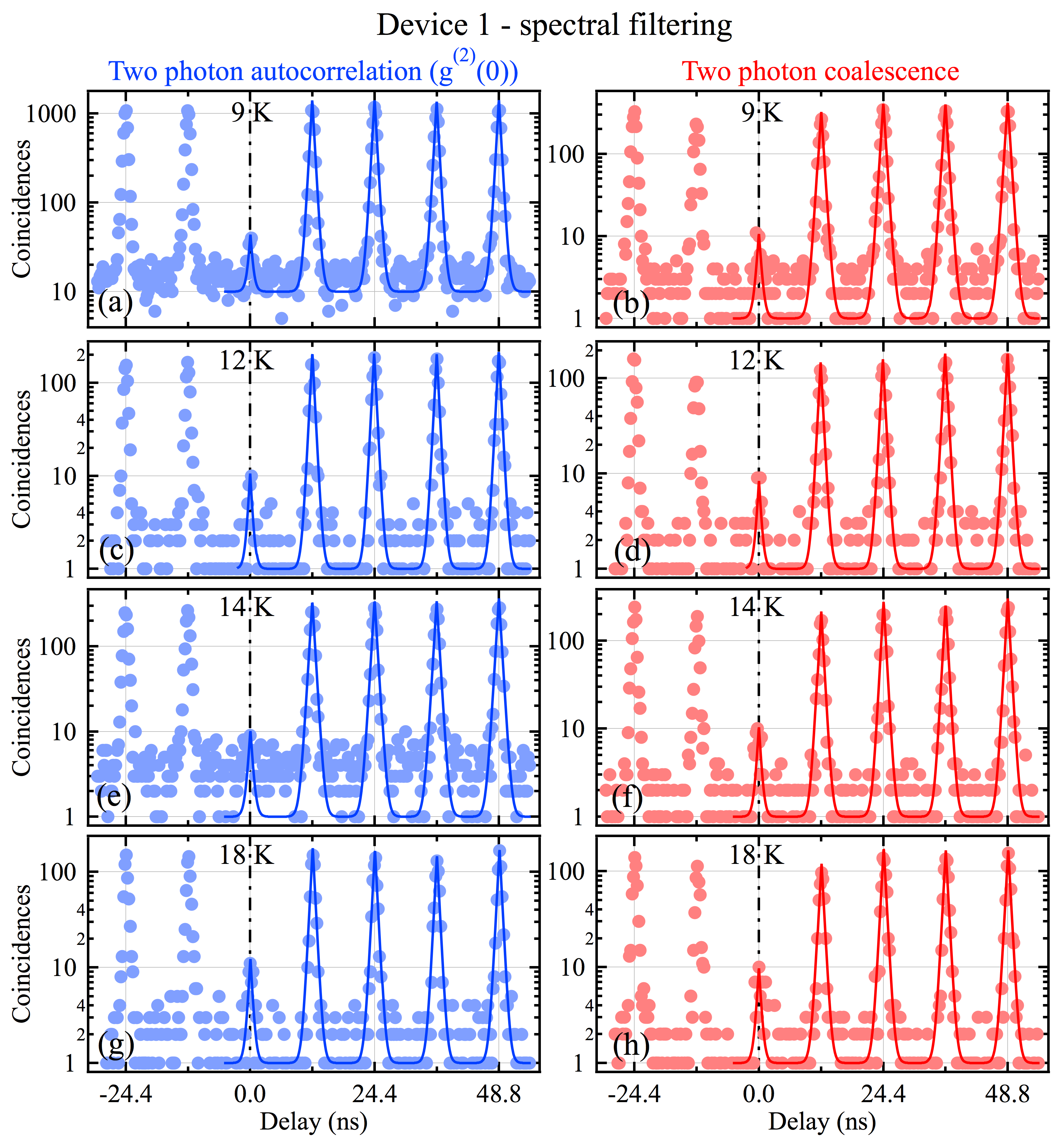} 
\end{centering}
\caption{\red{Two-photon coincidence histograms for Device 1 at different temperatures under resonant excitation with spectral filtering. Left (resp. right) column corresponds to the autocorrelation function of second order ($g^{(2)}(\tau)$) (resp. indistinguishability ($I$) of two photons delayed by 12.2 ns). Every row corresponds to a different temperature (from the top to the bottom): 9, 12, 14, 18 K, respectively. The vertical axis in every panel is shown in log scale. The full lines represent the optimal fitting to the experimental data points.}}
\label{Device1_9K_etal}
\end{figure}

\begin{figure}
\begin{centering}
\includegraphics[width=0.7\textwidth]{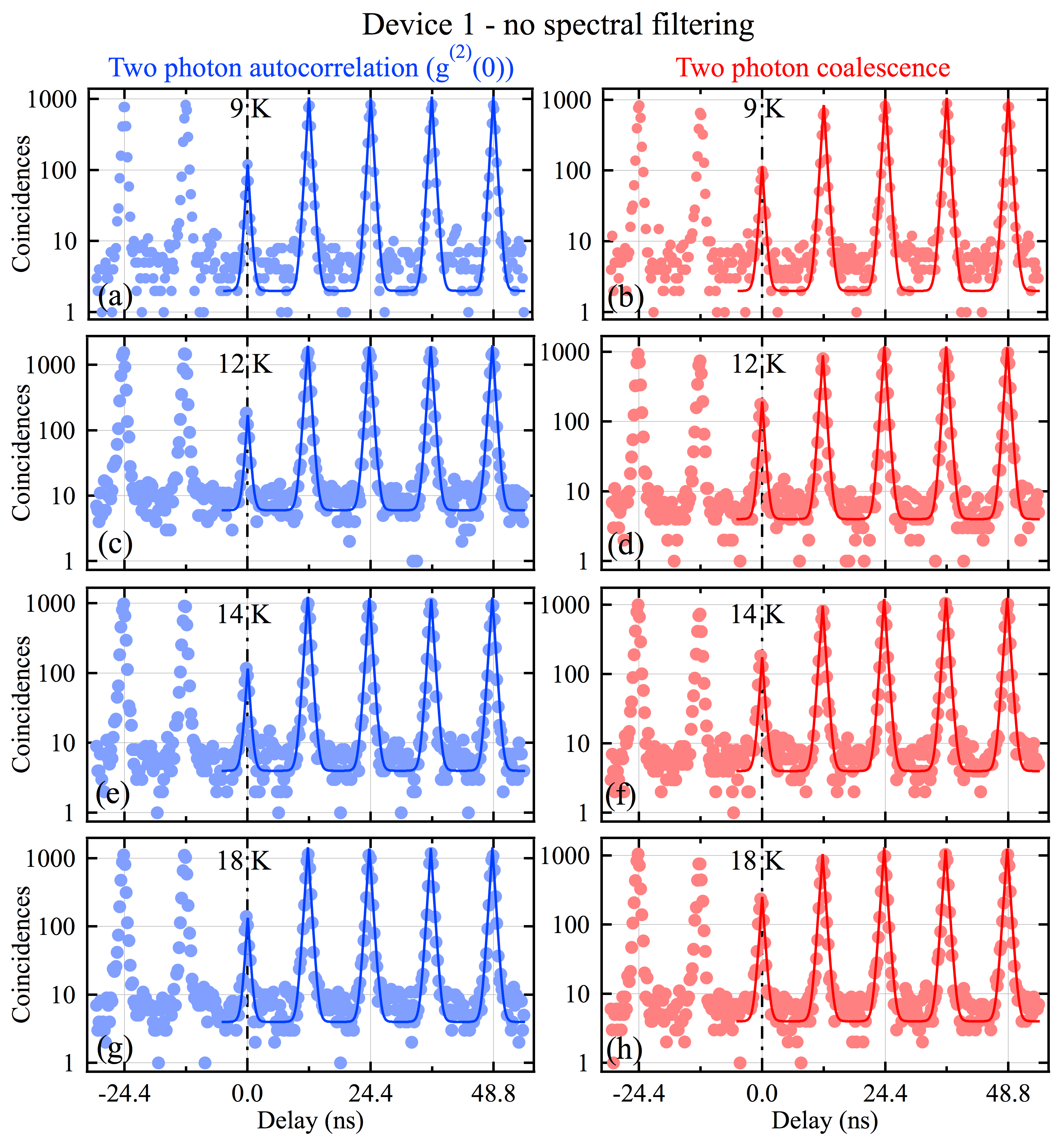} 
\end{centering}
\caption{\red{Two-photon coincidence histograms for Device 1 at different temperatures under resonant excitation without spectral filtering. Left (resp. right) column corresponds to the autocorrelation function of second order ($g^{(2)}(\tau)$) (resp. indistinguishability ($I$) of two photons delayed by 12.2 ns). Every row corresponds to a different temperature (from the top to the bottom): 9, 12, 14, 18 K, respectively. The vertical axis in every panel is shown in log scale. The full lines represent the optimal fitting to the experimental data points.}}
\label{Device1_9K_noetal}
\end{figure}

The experimental values obtained from the coincidence histograms of the second order autocorrelation function ($g^{(2)}(0)$) and the two photon coalescence are obtained in a standard Hanbury Brown and Twiss and a HOM setup, respectively. The emission signal is split using a first beam-splitter, and a 12.2 ns delay is implemented on one of the arms  to ensure that single photons generated in two consecutive excitation pulses temporally overlap on a fiber beamsplitter where the HOM measurement takes place. The coefficients of the HOM beamsplitter are $R = 0.502$ and $T = 0.498$. The output fibers of the beamsplitter are  connected to fibered SPADs, which are connected to a HydraHarp 400 autocorrelator to extract the two-photon coincidence histogram.

\red{Figure \ref{Device1_9K_etal} [\ref{Device1_9K_noetal}] presents, as an example, the histograms in logarithmic scale of $g^{(2)}(0)$ and indistinguishability (left and right column) of Device 1 at different temperatures (see the different rows corresponding to different temperatures from 9 to 18 K) with [without] spectral filtering. We fit every histogram, from -15 to 600 ns time delay, with a series of double-exponential decay functions (following the form $A_i e^{-\left|\frac{t-t_i}{\tau}\right|}$, with different amplitudes and temporal positions but a common temporal decay $\tau$) on top of a constant value accounting for the time independent background arising from the SPADs dark counts ($10^2-10^3$ counts/s)}. The impact of the fluctuations of these dark counts has been considered to establish the error bounds in both $g^{(2)}(0)$ and indistinguishability. For every histogram the best fit (see full line) is superimposed to the experimental data points (see full circles).

The ratio of the peak at zero delay ($A_0$) and the average area of 50 uncorrelated peaks ($\left\langle A\right\rangle$) renders the value of $g^{(2)}(0)$. The observed values for the $g^{(2)}(0)$ increase without the etalon filtering with respect to those obtained after spectral filtering, because of the increased residual resonant laser light scattered by the device (compare the $g^{(2)}(0)$ peaks on the left column in Fig. \ref{Device1_9K_noetal} with respect to those on the left column in Fig. \ref{Device1_9K_etal}). Since such residual laser light would be suppressed with a better polarization extinction ratio, the intrinsic photon indistinguishability is extracted by correcting from this residual $g^{(2)}(0)$.
The indistinguishability is given by:

\begin{equation}
I=\frac{1}{(1-\epsilon)^2}\left(g^{(2)}(0)+\frac{R^2+T^2}{2 R T}-\frac{(R+T)^2}{2 R T}\frac{A_0}{\left\langle A\right\rangle}\right)
\label{indis}
\end{equation}

where $(1-\epsilon)=0.995$ corresponds to the classical visibility of the interferometer (tested with a continuous wave laser at the wavelength of the QD emission) and the ratio of $A_0/\left\langle A\right\rangle$ corresponds to the area of the peak at zero delay and the average value of 49 uncorrelated peaks (excluding those at the temporal delays $\pm12.2$ ns). In Figs. \ref{Device1_9K_etal} and \ref{Device1_9K_noetal} we show all the histograms of the autocorrelation function of second order ($g^{(2)}(\tau)$) and the indistinguishability for the four measured temperatures (see Fig. 1(e) and 2 of the main text) with and without spectral filtering, respectively.

\begin{figure}
\begin{centering}
\includegraphics[width=0.50\textwidth]{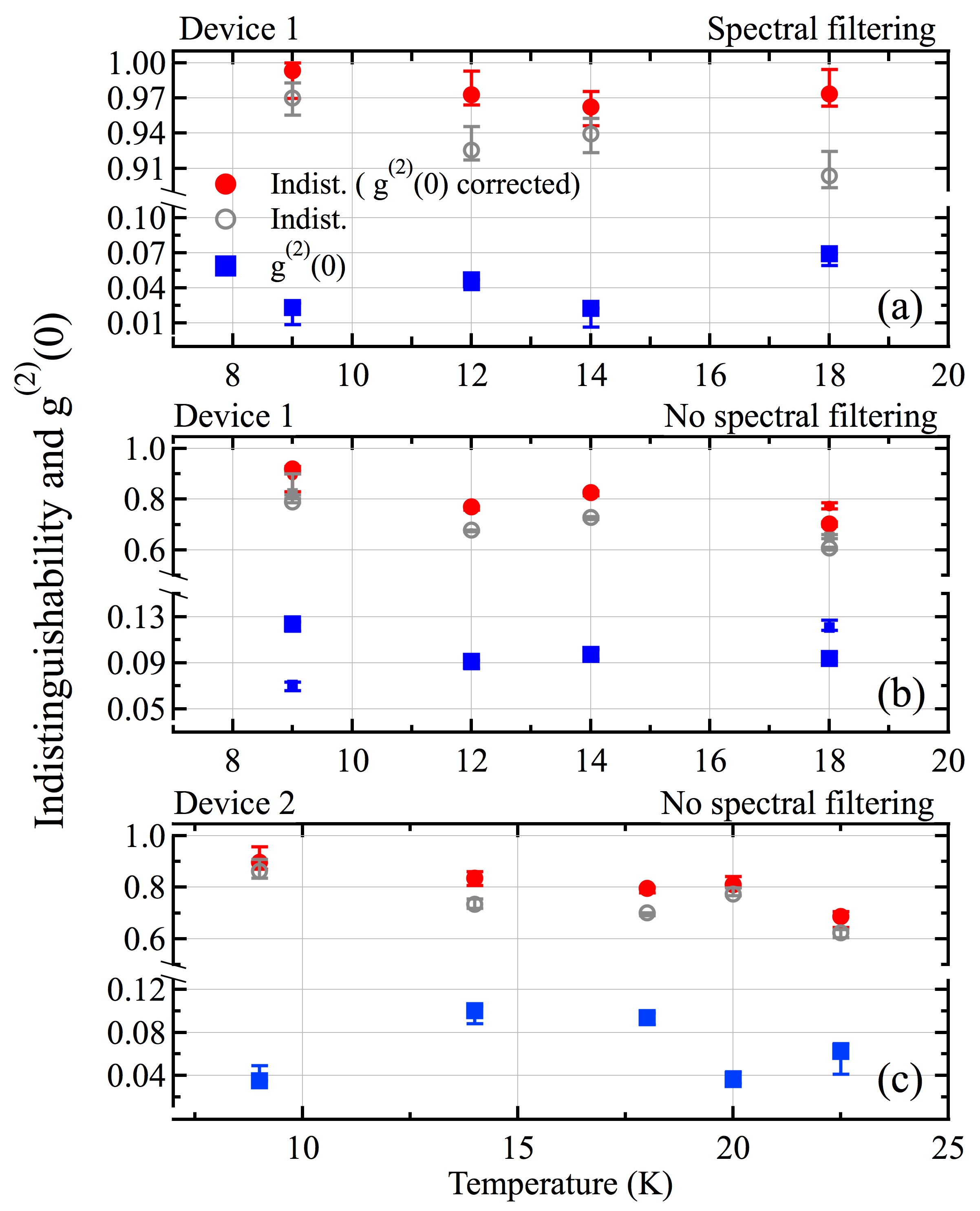} 
\end{centering}
\caption{Dependence of the indistinguishability and $g^{(2)}(0)$ as function of the temperature. Full red (open gray) circles represent the corrected (uncorrected) indistinguishability from the corresponding $g^{(2)}(0)$ value, in full blue squares. Panel (a)/(b) corresponds to Device 1 with (without) spectral filtering. Panel (c) corresponds to Device 2 without spectral filtering.}
\label{g2_M_vs_T}
\end{figure}

Figure \ref{g2_M_vs_T} compiles all the results from Devices 1 and 2 on the $g^{(2)}(0)$ (blue squares) and indistinguishability with and without the $g^{(2)}(0)$ correction described in Eq. \ref{indis}, see full red and open gray circles, respectively. Error bars in the $g^{(2)}(0)$ data points have been obtained by considering the error bar on the baseline accounting for the dark counts The error bars on the indistinguishability are dominated by the  error on the $g^{(2)}(0)$ and on the base line.

\section{Section 4: Theory}

\subsection{Coupling with acoustic phonons}

The interaction Hamiltonian between excitons and longitudinal acoustic (LA) phonons has the general form
\begin{equation}
H = \sum_{i,j} \vert \psi_i \rangle \langle \psi_j \vert \left[ \sum_q  M_q^{(ij)}(\hat{b}_q+ \hat{b}_q^{\dagger}) , \right]
\end{equation}
where $\vert \psi_i \rangle$ are exciton wave functions, and $\hat{b}_q^{(\dagger)}$ the ladder operators for the LA phonons. This interaction can be divided into two parts.
The diagonal part of this interaction (i.e. terms $i=j$) gives rise to phonon sidebands (PSBs) for each of the excitonic transition. The non-diagonal part is responsible for pure dephasing processes of the lowest excitonic transition through virtual transitions with higher excitonic levels \cite{muljarov04,grange2009decoherence}. As discussed in the paper, the latter is described through the temperature-dependent parameter $\gamma^*$ describing Markovian pure dephasing.

The PSB involves the diagonal coupling term for the fundamental exciton ($i=j=0$), with coupling form factors $M_q=M_q^{00}$ which read:
\begin{equation}
M_q = \sqrt{\frac{\hbar q}{2 c_s \rho_m V}}  \left[D_e \langle \phi_e | e^{-qr_e}| \phi_e \rangle + D_h \langle \phi_h |e^{-qr_h} | \phi_h \rangle \right]
\end{equation}
where $q$ is the phonon wavevector, $V$ the crystal volume, $|\phi_{e/h} \rangle$ the electron/hole envelope wavefunctions of the exciton, $D_e$ and $D_h$ the deformation potential for the electrons and holes respectively.
We take $c_s = 5110$~m.s$^{-1}$ for the speed of sound, $\rho_m = 5370$~kg.m$^{-3}$ for the mass density, following typical GaAs bulk values.
Identical envelope wavefunctions are assumed for electrons and holes with isotropic Gaussian shape:
\begin{equation}
\phi_{e/h}(r) =\frac{e^{-r^2/2\sigma^2}}{\pi^{3/4} \sigma^{3/2}} ,
\end{equation}
where $\sigma=5$~nm is the electron/hole confinement length.
This leads to a exciton-phonon coupling term of
\begin{equation}
M_q = D \sqrt{\frac{\hbar q}{2 c_s \rho_m V}}  e^{-\sigma^2 q^2/4} ,
\end{equation}
where $D=D_e+D_h$ is the exciton deformation potential. From the measured spectrum at 9~K we extract $D=14$~eV (see below), which is used throughout this work.

\begin{figure}
\begin{centering}
\includegraphics[width=0.45\textwidth]{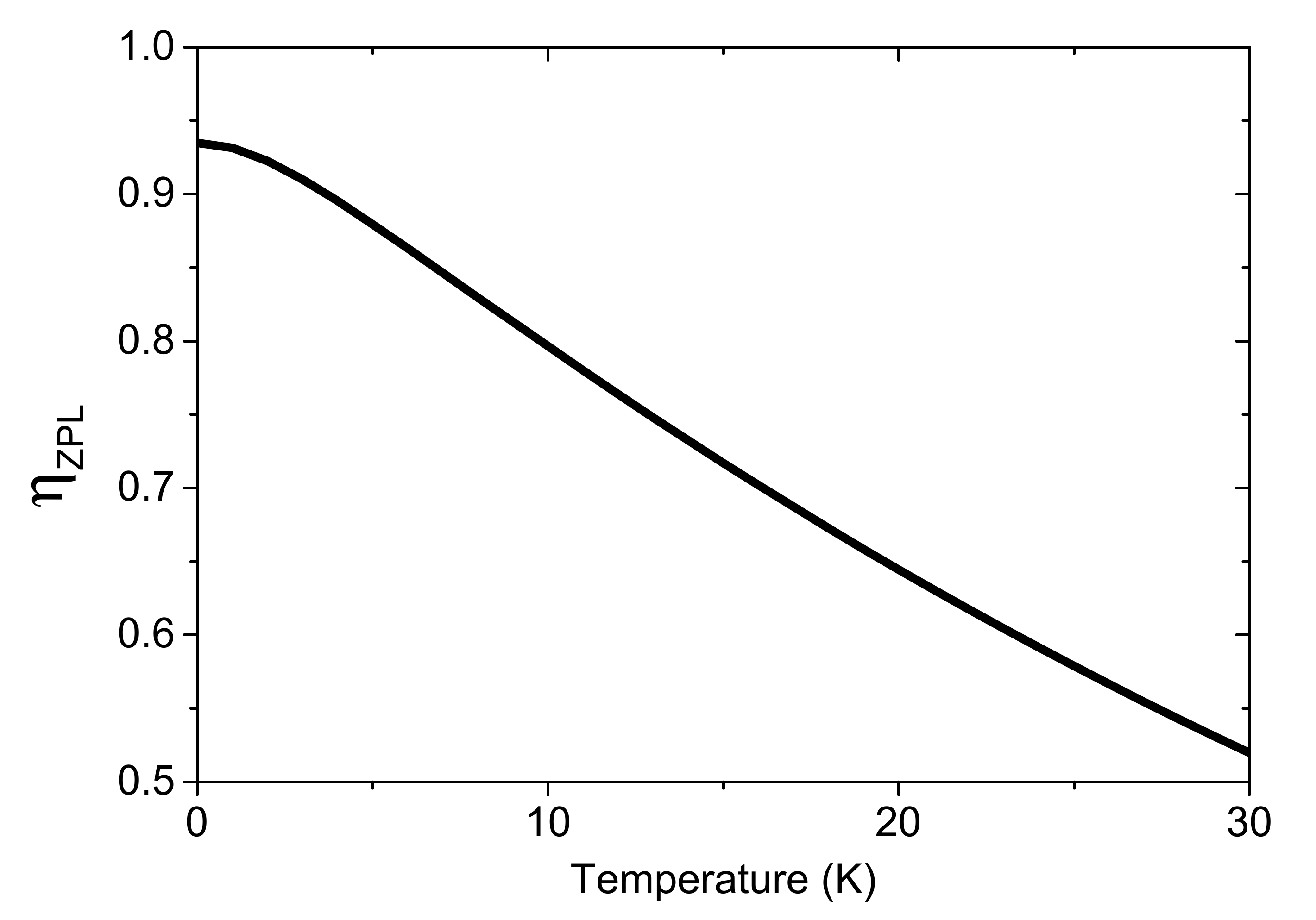} 
\end{centering}
\caption{ZPL fraction $\eta_{\text{ZPL}}$ as a function of temperature, in the bulk case.
}
\label{eta_zpl_bulk}
\end{figure}

In the bulk case (i.e. no cavity-QED effect), the non-Markovian dephasing dynamics due to the PSBs is provided by the exact solution of the independent boson model \cite{mahan2000many}. Following a QD excitation at $t=0$, the QD polarization reads for $\tau \ge 0$:
\begin{equation}
\langle \hat{c}^{\dagger}(t+\tau)\hat{c}(t)\rangle = \text{exp}\left[-i \frac{\gamma}{2} t -i \frac{\gamma^*}{2} (t+\tau) - \int d\omega \frac{J(\omega)}{\omega_q^2} \left( N(\omega)[1-e^{i\omega\tau}]+[N(\omega)+1][1-e^{-i\omega\tau}] \right) \right]
\end{equation}
where $\hat{c}^{\dagger}$($\hat{c}$) are the exciton creation (annihilation) operators. The first two terms in the exponential are the Markovian decay term for populations and coherence respectively, while the integral describes the phonon sidebands. $J$ is the spectral function of the phonon bath:
\begin{equation}
J(\omega) = \sum_q M_q^2 \delta(\hbar\omega - \hbar\omega_q) = \frac{\omega^3}{4\pi^2  \rho_m c_s^5  }
D^2 e^{-\sigma^2 q^2/2}.
\end{equation}
The corresponding emission spectrum can be expressed as:
\begin{equation}
  S_{\text{QD}}(\omega-\omega_{\text{0}}) = \int_{-\infty}^{\infty} \dd \tau \exp\left[ -\frac{(\gamma + \gamma^*)}{2} |\tau| - \int_{-\infty}^{\infty} \!\dd\omega\, \frac{J(\omega)}{\omega_q^2} (N(|\omega|)+\theta(\omega))\left( 1 - e^{-i\omega\tau} \right) \right] e^{i\omega\tau}
 \label{SQD}
\end{equation}
$\theta(\omega)$ being the Heaviside step function. The zero-phonon line (ZPL) consists of a Lorentizian of full width at half maximum (FWHM) $\gamma + \gamma^*$ and weight
\begin{equation}
\eta_{\text{ZPL}} = \int_0^{\infty} \!\dd\omega\, \frac{J(\omega)}{\omega_q^2} (2N(\omega)+1) .
 \label{etaZPL}
\end{equation}
The temperature dependence of $\eta_{\text{ZPL}}$ is indicated in Fig.~\ref{eta_zpl_bulk}.

In the cavity-QED case, the properties of the emitted photons are calculated using the non-equilibrium Green's function (NEGF) approach presented in Ref.~\cite{hornecker2016influence}. In this approach, a polaron transformation is used, which provides an exact solution of the problem if the QD-cavity interaction is turned off.  Within the polaron frame, the QD-cavity exchange terms describing the zero-phonon line (ZPL) are treated exactly. On the other hand, the phonon-assisted exchange terms between the QD and the cavity are treated within the self-consistent Born approximation.
An instantaneous excitation of the QD at $t=0$ is assumed. Coupling of strength $g$ to a single cavity mode of linewidth $\kappa$  is considered. The other radiative channels are treated by a decay rate $\gamma$, while pure dephasing is described through the constant $\gamma^*$. The two-time correlators $\langle \hat{a}^{\dagger}(t_2)\hat{a}(t_1)\rangle$ of the cavity field are computed by solving the Kadanoff-Baym equations.
The indistinguishability, which represents the probability of two-photon interference assuming purely single photons, is then calculated using \cite{kiraz2004quantum,grange2015cavity}
\begin{equation}
I=  \frac{\int_{0}^{\infty} dt \int_0^{\infty} d\tau \vert\langle \hat{a}^{\dagger}(t+\tau)\hat{a}(t)\rangle\vert^2}{\int_{0}^{\infty} dt \int_0^{\infty} d \tau \langle \hat{a}^{\dagger}(t)\hat{a}(t)\rangle\langle \hat{a}^{\dagger}(t+\tau)\hat{a}(t+\tau)\rangle} .
\end{equation}

\subsection{Effective Purcell factor}

We note $\gamma_0$ the radiative decay rate of the QD in the bulk case (i.e. without cavity effect). For a QD coupled to the cavity mode, the total decay rate becomes
\begin{equation}
\gamma = \gamma_0 + \gamma_0 F_{\text{eff}}  .
\end{equation}
The first term represents the radiative rate outside of the cavity mode, which equals the one of the bulk in our microcavity system. The second represents the emission rate into the cavity mode. The effective Purcell factor $F_{\text{eff}}$ is hence defined as the ratio of the emission rate into the cavity over the bulk one.
In absence of PSBs, the emission rate into the cavity mode is given by \cite{auffeves2009pure}
\begin{equation}
\gamma_0 F_{\text{eff}}^{\text{no phonon}} = \frac{4g^2}{\kappa+\gamma_0+\gamma^*} \frac{1}{1+\left( \frac{2\delta}{\kappa+\gamma_0+\gamma^*} \right)^2} ,
\end{equation}
where $g$ is the QD-cavity coupling strength, $\kappa$ the cavity linewidth, and $\delta = \omega_{\text{QD}}-\omega_{\text{cav}}$ is the detuning between the exciton and the cavity mode.
The above formula is valid in the weak QD-regime coupling, if the cavity linewidth is much larger than the QD linewidth (i.e. the so-called bad-cavity regime).
Note that the fraction of emission into the cavity mode $\beta$ is given by
\begin{equation}
\beta = \frac{F_{\text{eff}}}{F_{\text{eff}}+1} .
\end{equation}
Conversely, $F_{\text{eff}}$ can be computed from $\beta$, as $F_{\text{eff}}= \beta/(1-\beta)$. This lattest approach is used to compute it from the dynamics calculated with the NEGF model.

In presence of phonon sidebands (PSBs), if the cavity is close to resonance with the ZPL, the cavity has primarily a Purcell effect on (i) the ZPL and (ii) the PSBs that match the cavity spectrum. The PSBs that fall outside the cavity spectrum do not emit in the cavity mode. The effective Purcell $F_{\text{eff}}$ is hence reduced compared to the no-phonon case. In the limit of the bad-cavity weak coupling regime, a good approximation is provided by the following formula:
\begin{equation}
\gamma_0 F_{\text{eff}}^{\text{no phonon}} = \eta_{\text{ZPL}} \frac{4  g^2}{\kappa+\gamma_0+\gamma^*} \frac{1}{1+\left( \frac{2\delta}{\kappa+\gamma_0+\gamma^*} \right)^2} + 2 \pi g^2 \int d \omega\rho_{\text{PSB}}(\omega) S_{\text{cav}}(\omega) ,
\end{equation}
where $\rho_{\text{PSB}}$ is the density of state in the PSBs, verifying the normalization $\int d\omega \rho_{\text{PSB}}(\omega) = 1-\eta_{\text{ZPL}}$. The first term describes the radiative rate of the ZPL, while the second one describes the radiative rate of the PSB emitted in the cavity.

\subsection{Accounting for the cavity mode splitting}

As sketched in Fig.~\ref{scheme_excitation}, the cavity mode in which the excitation is made (noted E) is assumed to be perfectly on resonance with the QD transition. The orthogonal cavity mode in which the signal is monitored (noted M) is hence detuned by $\delta_{\text{EM}}$ from both the exciting cavity mode and the QD transition.
The QD transition is assumed to be equally coupled to the two cavity modes (with a coupling strength $g/\sqrt{2}$). The sum of the radiative rates that are not collected in the M cavity mode are:
\begin{equation}
\gamma
 = \gamma_0 + \eta_{\text{ZPL}} \frac{4   \frac{g^2}{2}}{\kappa+\gamma_0+\gamma^*} + 2 \pi \frac{g^2}{2} \int d \omega\rho_{\text{PSB}}(\omega) S_{\text{E}}(\omega)
\end{equation}
Such Markovian decay rate for lossy channels is taken as an input in the NEGF calculation, assuming a detuning of $\delta_{\text{EM}}$ between the QD transition and the M-cavity mode, and a coupling strength $g/\sqrt{2}$.
The parameters extracted from the experiments are $\delta_{\text{EM}}= \omega_M - \omega_E = 80~\micro$eV for Device~1, while $\delta_{\text{EM}}= \omega_M - \omega_E = -40~\micro$eV for Device~2.
Finally, all the parameters used for the two devices are summarized in Table~\ref{parameters}.

\begin{table}
\begin{tabular}{c|c|c|c|c|c|c}
\hline
Device& QD-cavity  & Cavity  & Nominal Purcell & Cavity mode  & Bulk decay & Parameters for  \\
& coupling ($\hbar g$) & linewidth ($\hbar \kappa$) & factor ($4g^2/(\kappa \gamma_ 0)$) & splitting $\hbar(\omega_{\text{M}}-\omega_{\text{E}})$  & rate ($\gamma_0$) & phonon sidebands\\
\hline
&&&&&&\\
Device~1 &  $19~\micro$eV & $90~ \micro$eV  & $\sim 24$ & $+80~ \micro$eV & 1 ns$^{-1}$ & $D=14~$eV, $\sigma = 5~$nm\\
&&  & & & & $c_s = 5110$~m.s$^{-1}$, $\rho_m = 5370$~kg.m$^{-3}$ \\
&&&&&&\\
Device~2 & $12~ \micro$eV &  $110~ \micro$eV & $\sim 8$ & $-40~ \micro$eV & idem & idem\\
&&&&&&\\
\hline
\end{tabular}
\caption
{Parameters used in the calculations for the two QD-cavity devices. In addition, identical temperature-dependent dephasing rate $\gamma^*$ is used for the two devices (see main paper).}
\label{parameters}
\end{table}

\section{Section 5: Additional results and discussion}

\subsection{Emission spectrum of the cavity-QD system: comparison between measurement and simulation}

\begin{figure}[h]
\begin{centering}
\includegraphics[width=0.90\textwidth]{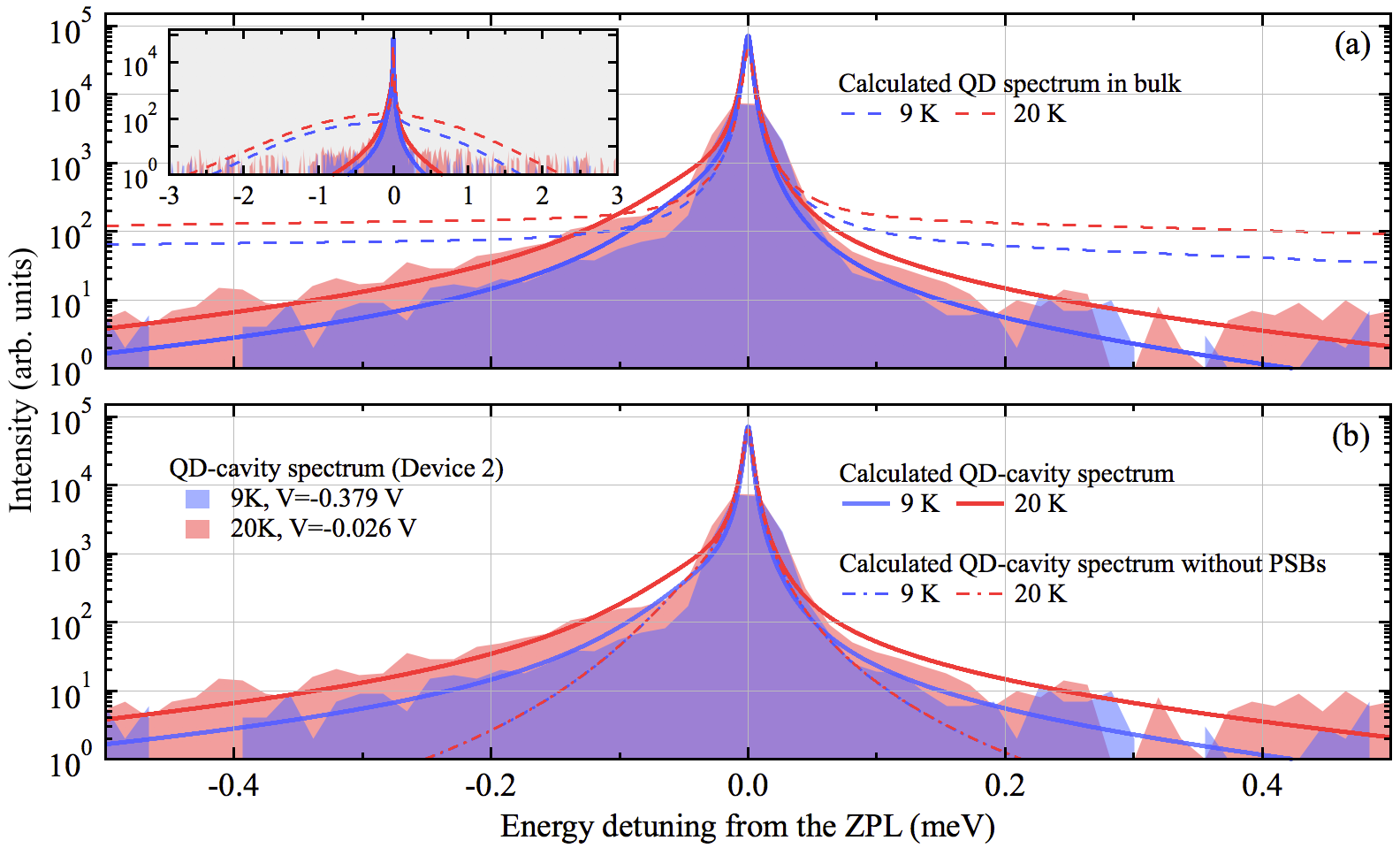} 
\end{centering}
\caption{\red{Measured (semitransparent filled blue/red areas) and calculated (full blue/red red lines) spectra at 9~K and 20~K for Device 2, same plot in both (a,b) panels. (a) The measurement and full calculations are compared to the calculated spectra of the QD in bulk (i.e. without any cavity-QED effect) at 9~K and 20~K (dashed blue/red lines, respectively). Inset: same plot on a larger energy scale, showing the strong reduction of phonon sidebands thanks to cavity-QED. (b)  The measurement and full calculations are compared to the calculated spectra without coupling to phonons  (dot-dashed blue/red lines). The comparison shows the residual contribution of phonon sidebands in the QD-cavity spectrum.}}
\label{spectra}
\end{figure}

\red{The measured spectra for the Device~2 at 9~K and 20~K are shown in Fig.~\ref{spectra}, along with the calculated one. In Fig.~\ref{spectra}(a), measurements and full calculations are compared to the calculated spectra without cavity-QED effect. This comparison shows the strong reduction of phonons sidebands (PSBs) thanks to cavity-QED, most visible in the figure insert. This effect is all the more important as the temperature increases. As discussed in the main paper, this cavity-QED redirection of the PSBs into the ZPL allows for a strong enhancement of indistinguishability.}

\red{In Fig.~\ref{spectra}(b), the same measurements and full calculations are compared  to the calculated spectra without considering PSBs. The deviation between the calculations with and without considering the coupling to phonons shows the residual presence of PSBs in the cavity-QED case. The influence of these residual PSBs on the indistinguishability is discussed below.}

\subsection{Relative influence of phonon sidebands and pure dephasing on indistinguishability}

\begin{figure}
\begin{centering}
\includegraphics[width=0.6\textwidth]{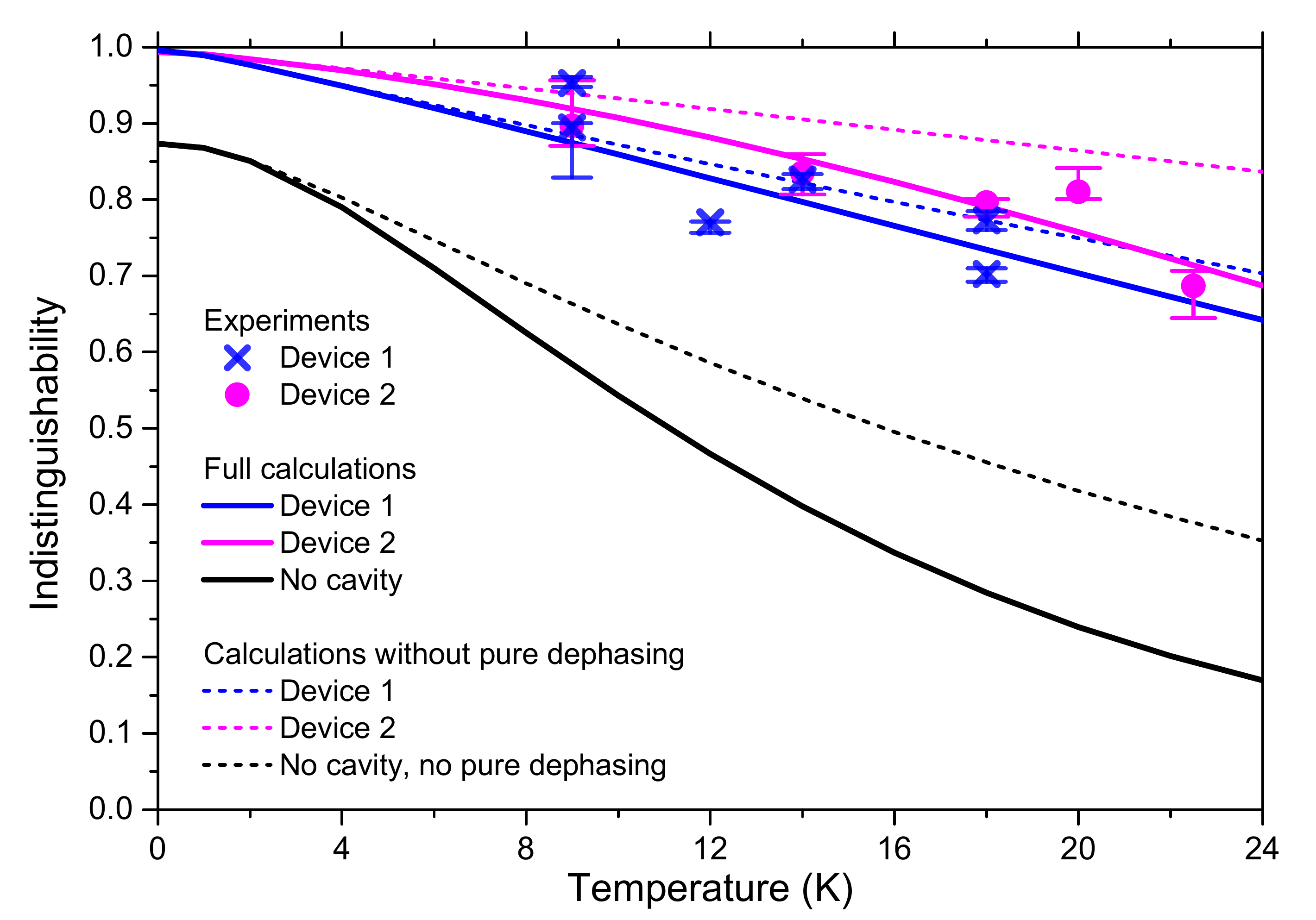} 
\end{centering}
\caption{Same figure as Fig.~2 of the paper, with additional dashed lines representing the calculations of indistinguishability vs temperature for the two devices assuming no pure dephasing (i.e $\gamma^*=0$). Hence these dashed lines represent the sole effect of the residual PSBs on the indistinguishability, while the ratio between dashed and solid lines represent the effect of pure dephasing only.
}
\label{IvsT_supp}
\end{figure}

\begin{figure}
\begin{centering}
\includegraphics[width=0.55\textwidth]{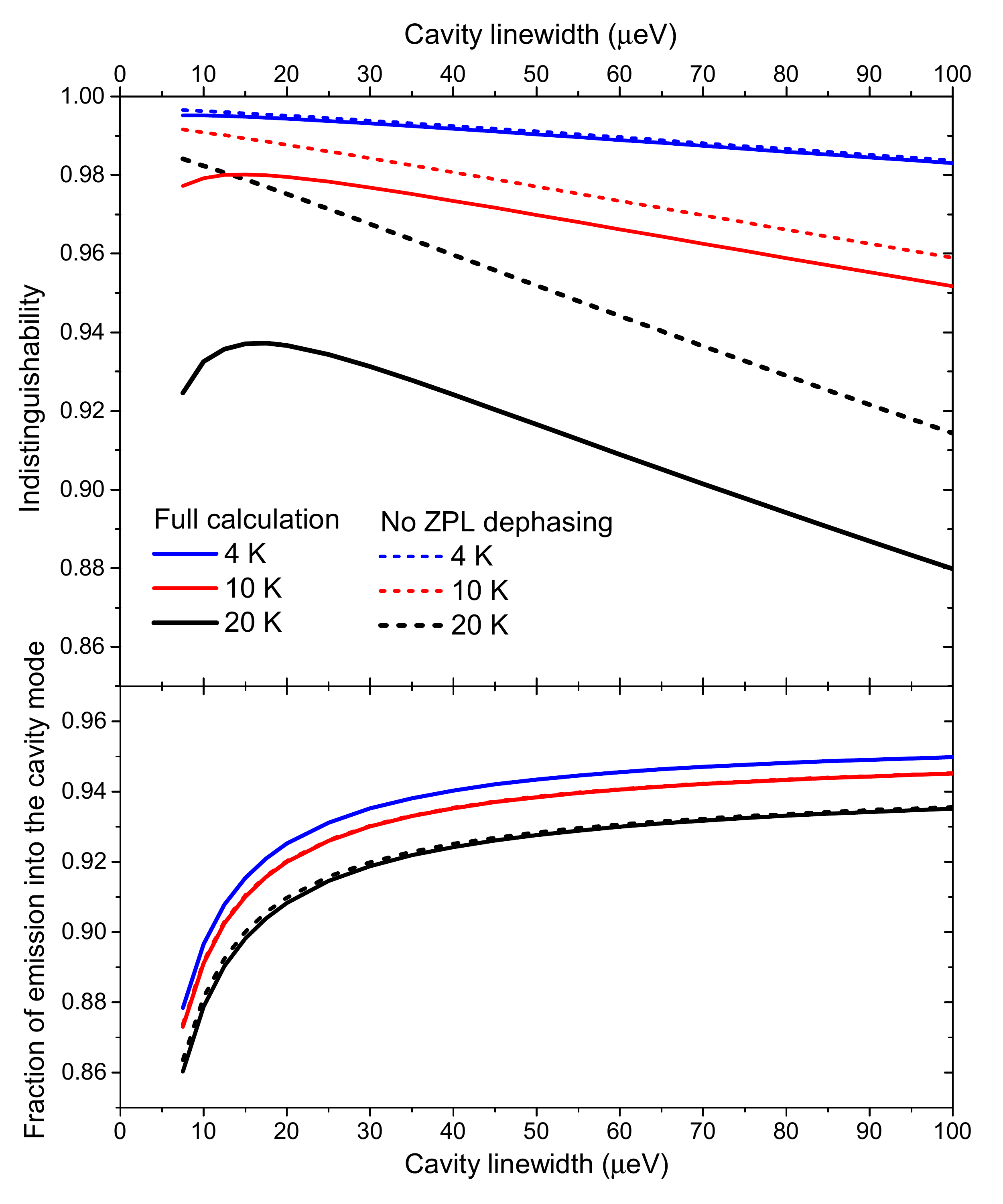} 
\end{centering}
\caption{
Same figure as Fig.~3 of the paper, with additional dashed lines representing the calculations of indistinguishability assuming no pure dephasing (i.e $\gamma^*=0$).
}
\label{Ivskappa_supp}
\end{figure}

\begin{figure}
\begin{centering}
\includegraphics[width=0.6\textwidth]{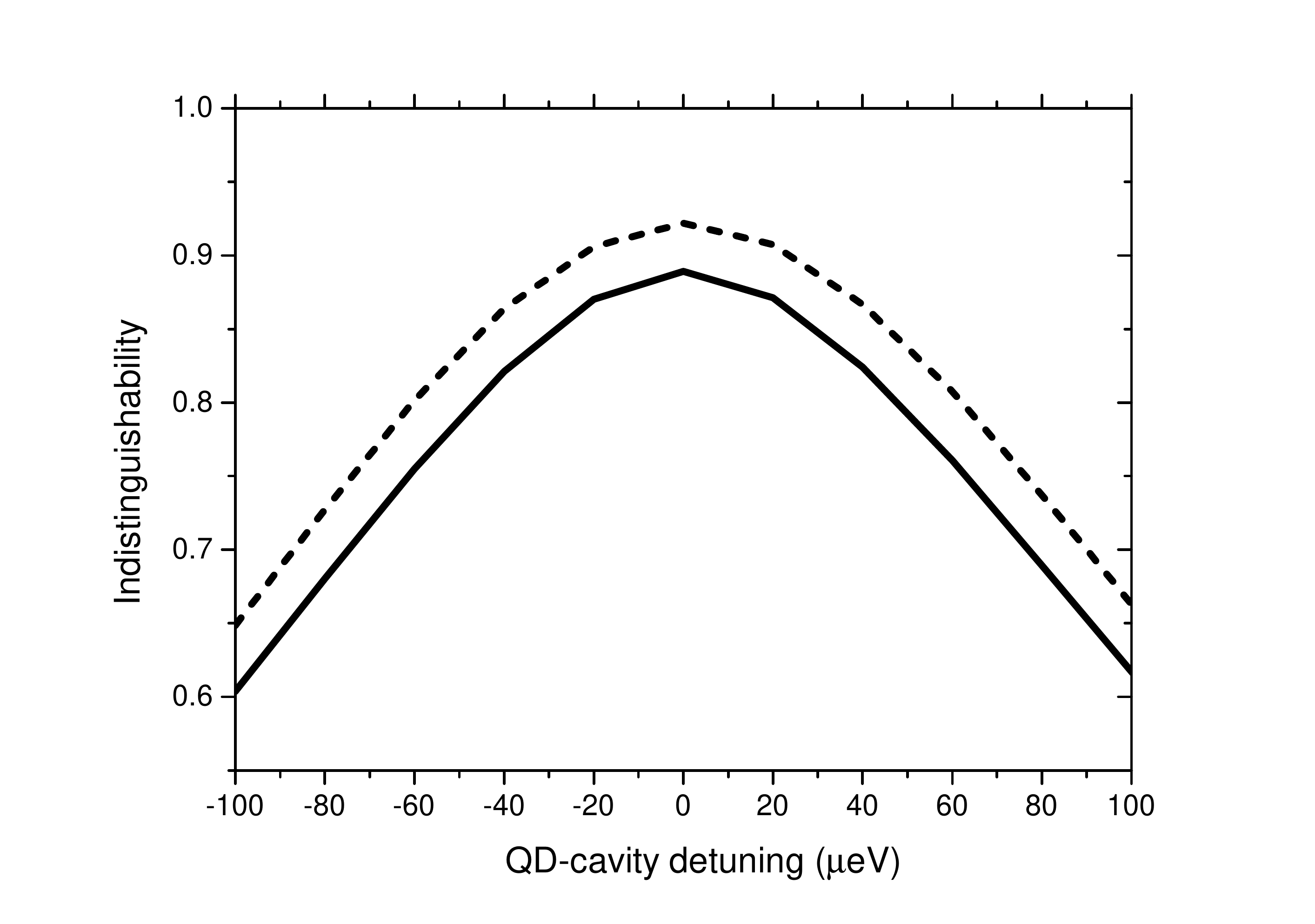} 
\end{centering}
\caption{Indistinguishability of the photons emitted by the QD-cavity system as a function of the detuning between the QD transition and the cavity mode for a temperature of 20~K. Full calculation (solid line) and calculation without pure dephasing (dashed line) are shown. No cavity mode splitting is considered. The other parameters are identical to those of Device~1.
}
\label{IvsDelta}
\end{figure}

In Fig.~2 and Fig.~3 of the paper, the calculations include both effects of PSBs and pure dephasing. Here we show and discuss the relative influence of these two effects. To this aim, we show in Fig.~\ref{IvsT_supp} and Fig.~\ref{Ivskappa_supp} the same figures as Fig.~2 and Fig.~3 of the paper, with additional calculations (dashed lines) assuming no pure dephasing (i.e. $\gamma^*=0$). Hence these calculations in which the pure dephasing is artificially turned off show the sole influence of the PSBs on the indistinguishability. On the other hand, the ratio between the calculations without dephasing (dashed lines) and the full calculations (solid lines) represents the effect of pure dephasing.

In Fig.~\ref{IvsT_supp}, we remark that a slightly more robust indistinguishability is obtained for Device~2 with respect to Device~1.
Yet, Device~1 has a larger effective Purcell (11.5 at 10~K)  factor than Device~2 (5.3 at 10~K), and thus better overcomes the effect of pure dephasing, as seen by the smaller ratio between the dashed and solid lines.
However due to a larger cavity splitting, the effect of PSBs is larger in Device~1 than in Device~2. Indeed, $\delta_{\text{EM}}= \omega_M - \omega_E = 80~\micro$eV for Device~1, while $\delta_{\text{EM}}= \omega_M - \omega_E = -40~\micro$eV for Device~2.
The larger detuning in Device~1 results in a larger fraction of the emission in the PSB with respect to the ZPL.
Yet for vanishing temperatures, the blueshift of the Device 1 cavity is more favorable as phonon absorption is not possible. This explains the crossing of the two curves for $T \sim 1$~K.

Fig.~\ref{IvsDelta} shows the indistinguishability as a function of the detuning between the QD transition and the cavity mode for a temperature of 20~K. As the detuning increases, the cavity enhancement of the ZPL decreases. In contrast, the emission rate in the PSBs stays almost constant for such detunings smaller than its width. As a consequence, the ZPL fraction decreases (i.e. the PSB fraction increases), so that the indistinguishability decreases. In this figure, the two cavity modes are assumed to be degenerate. However, a similar dependence is obtained if one considers the E cavity mode (i.e. in which the excitation is made) to be always resonant with the QD transition.

Fig.~\ref{Ivskappa_supp} shows the same as in Fig.~3 of the paper, with additional dashed lines indicating the calculation without pure dephasing (i.e $\gamma^*=0$). As above (Fig.~\ref{IvsT_supp}), this allows to see the relative influence of the PSBs and pure dephasing. At 4~K, the effect of PSBs is dominant as the two curves overlap. With increasing temperature, the effect of pure dephasing, which corresponds to the ratio between dashed and solid lines, has a more important role. In particular, as the QD-cavity strong coupling is reached for small cavity linewidths, the emission speed decreases, as it becomes limited by the cavity lifetime. As a consequence, pure dephasing has a larger contribution with decreasing linewidth.

\subsection{Influence of the Purcell factor in the strategy of cavity-linewidth reduction}

\begin{figure}
\begin{centering}
\includegraphics[width=0.6\textwidth]{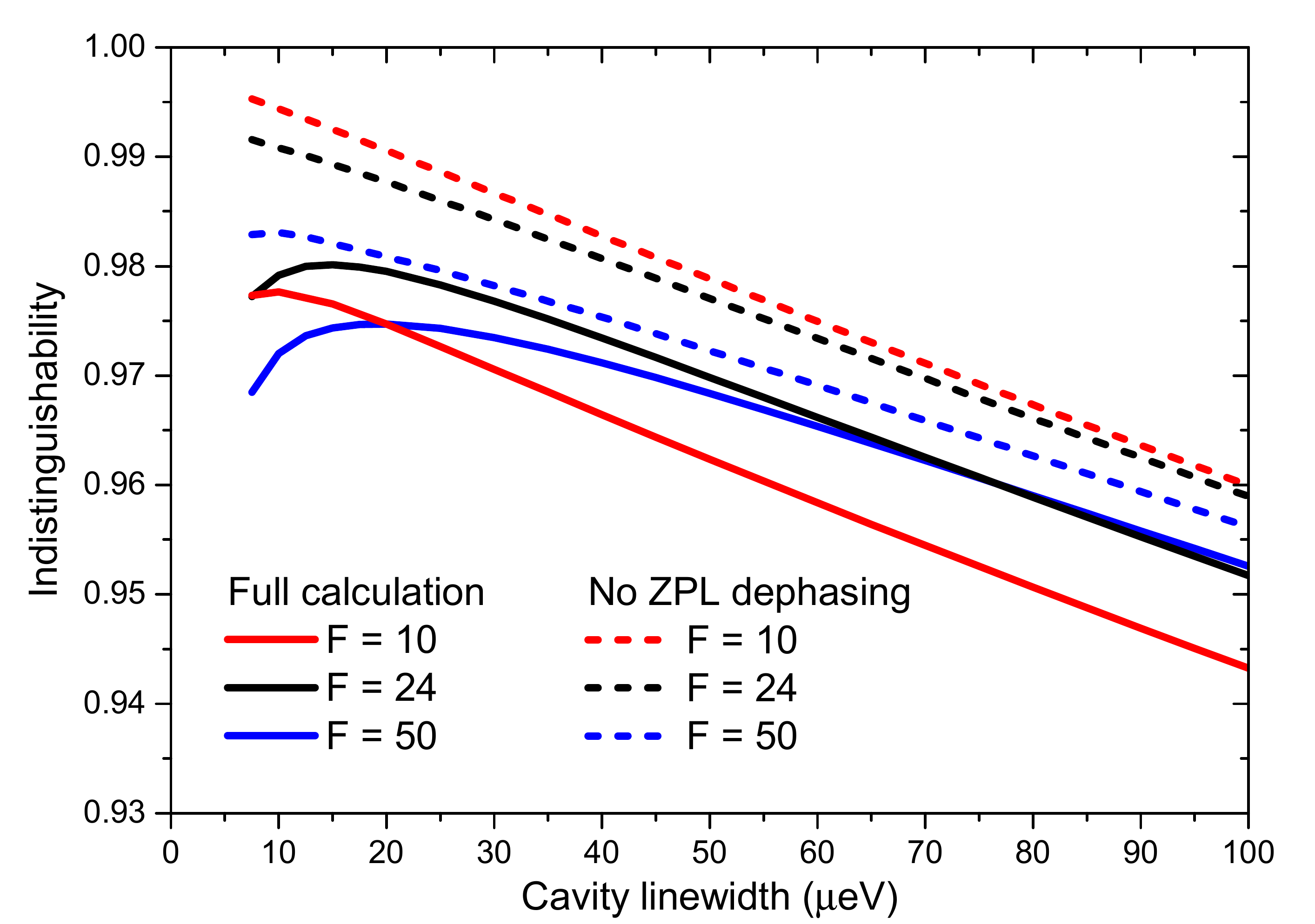} 
\end{centering}
\caption{Indistinguishability as a function of the Purcell factor for various nominal Purcell factors $F=4g^2/\kappa = $ 10, 24 and 50. Full calculations (solid lines) and calculations without pure dephasing (dashed lines) are shown.
}
\label{I_vs_kappa_10K_Purcell}
\end{figure}

Fig.~\ref{I_vs_kappa_10K_Purcell} shows the indistinguishability as a function of cavity linewidth for various nominal Purcell factors at a temperature of 10~K.
The full calculations (solid lines) as well as the calculations without pure dephasing (dashed lines) are shown.
We start by describing the calculations without pure dephasing considered (dashed lines).
For a fixed cavity linewidth, the indistinguishability decreases with increasing Purcell factor (and hence QD-cavity coupling strength). This effect is all more important as the QD-cavity strong coupling regime is approached, i.e. for small cavity linewidths. This is in agreement with the findings of Kaer et al \cite{kaer2013microscopic}.
If we now look at a given Purcell factor, the indistinguishability monotonously increases with decreasing linewidth. This effect is due to the reduction of the PSB fraction in the cavity-mediated emission. This former effect constitutes the core idea that we propose to exploit for further cavity optimization.

We now discuss the full calculations (solid lines), incorporating the effect of pure dephasing. The larger the Purcell factor, the more the effect of pure dephasing is reduced. Hence a compromise in terms of Purcell factor appears to be necessary considering these two effects, namely (i) the cavity suppression of PSBs and (ii) the overcoming of pure dephasing.
In conclusion a nominal Purcell factor of the order of 20 appears to be already optimal at 10~K for the proposed strategy of cavity linewidth reduction. Besides, this strategy is predicted to be valid in a large range of Purcell factors. Note that a low temperatures, lower Purcell factors will give the best indistinguishabilities as the effect of pure dephasing is suppressed.

\end{document}